\documentstyle[preprint,aps]{revtex}

\newcommand{\nc}{\newcommand}
\nc{\al}{\alpha}
\nc{\be}{\beta}
\nc{\ga}{\gamma}
\nc{\de}{\delta}
\nc{\ep}{\epsilon}
\nc{\ze}{\zeta}
\nc{\et}{\eta}
\nc{\ka}{\kappa}
\nc{\la}{\lambda}
\nc{\rh}{\rho}
\nc{\si}{\sigma}
\nc{\ta}{\tau}
\nc{\up}{\upsilon}
\nc{\ph}{\phi}
\nc{\ch}{\chi}
\nc{\ps}{\psi}
\nc{\om}{\omega}
\nc{\Ga}{\Gamma}
\nc{\De}{\Delta}
\nc{\La}{\Lambda}
\nc{\Si}{\Sigma}
\nc{\Up}{\Upsilon}
\nc{\Ph}{\Phi}
\nc{\Ps}{\Psi}
\nc{\Om}{\Omega}
\nc{\ptl}{\partial}
\nc{\del}{\nabla}
\renewcommand{\be}{\begin{eqnarray}}
\nc{\ee}{\end{eqnarray}}

\begin{document}
\title{A Non-associative Deformation of Yang-Mills Gauge Theory}
\draft
\preprint{UM-P-95/69; RCHEP-95/18}
\author{A. Ritz\footnote{ritz@physics.unimelb.edu.au}
and G. C. Joshi\footnote{joshi@physics.unimelb.edu.au}}
\address{Research Centre for High Energy Physics,\\
	School of Physics, The University of Melbourne, \\
	Parkville, VIC 3052 Australia}
\date{25$^{th}$ August 1995}

\maketitle

\begin{abstract}
 An ansatz is presented for a possible non-associative deformation
 of the standard Yang-Mills
 type gauge theories. An explicit algebraic structure for the
 deformed gauge symmetry
 is put forward and the resulting gauge theory developed.
 The non-associative deformation is constructed in such a way that an
 apparently associative Lie algebraic structure is retained modulo a closure
 problem for the generators.
 It is this
 failure to close which leads to new physics in the model
 as manifest in the gauge field kinetic term in the resulting Lagrangian. A
 possible connection between this model and quantum group gauge theories is
 also investigated.
\end{abstract}

\newpage
\section{Introduction}
Recently there has been considerable interest in the construction of
{\em quantum group gauge theories\/} (QGGTs)
\cite{arefeva91a,isaev92,hira92,castell92,arefeva93a,brzez93a,brzez93b,isaev93,arefeva93b,zhong94,castell94},
where a quantum
group plays the role of the gauge group. One source of motivation for this work
has been the suggestion that relaxing the rigid structure of the Lie group
based gauge theories may lead to new explanations for fundamental theoretical
problems such as spontaneous symmetry breaking and quark confinement.

Lie group theory, as the predominant mathematical tool for the analysis of
physical symmetries, has been extraordinarily successful in providing a
unified description of many aspects of particle physics. In particular the
currently accepted descriptions of the strong, weak,
and electromagnetic interactions,
have the common theme of an underlying gauge symmetry described by a Lie
group. Despite this success there continue to be difficulties in unifying
the various forces, and indeed in explaining some of the features
of the individual forces such as those mentioned above. It would seem
reasonable
to consider the possibility that a full unification of the fundamental forces
may require a mathematical structure beyond groups.
In the first instance we might intuitively expect that such a structure
would be generalisation of the Lie group.
The quantum group approach, regardless of how it was originally conceived,
is clearly a construction
of this type wherein the Lie group
gauge
symmetry is replaced by a more general quantum group symmetry which reduces
to the standard case in the limit of some parameter.

The transition from theories based on Lie group
internal symmetry spaces to those
where the symmetry is that of a quantum group has, however, proved to be rather
problematic. The main efforts have focussed on keeping the classical form of
the
gauge transformations. The gauge potential $A$ transforms as follows,
\be
 A & \longrightarrow & A' = UAU^{-1} - \frac{i}{g} (\ptl U)U^{-1},
\ee
where $U$ is chosen to be an element of a quantum
group.
The difficulty as described by Aref'eva and Arutyunov \cite{arefeva93b}
is to determine the relevant differential
calculus and also the algebra from which the gauge potentials $A$
should be drawn to ensure that $A'$ also belongs to that algebra.
Recently it has
been claimed \cite{arefeva93b} that it is only possible to present an algebraic
group gauge potential based on $U_q(N)$, and that groups such as $SU_q(N)$ are
not allowed. It has also been claimed \cite{brzez93a} that if the gauge fields
have values in $U_q(g)$, the quantum universal enveloping algebra
of the Lie algebra $g$, then the resulting gauge theory will be
isomorphic to the non-deformed theory if the base space is classical spacetime.
The implication being that to obtain non-trivial
results an underlying quantum space must be considered
\cite{brzez93b}. Although the situation is far from clear at this stage it
would
appear that the most general QGGTs require detailed analysis of both the
differential calculus on quantum groups (see for example Woronowicz
\cite{woron87,woron89}),
and also the non-commutative geometric structure of quantum spaces.

With this complex situation in mind, we present in this
letter an alternative approach
to the generalisation of the standard Yang-Mills type gauge theories.
Our approach will be to extend the standard (Lie) gauge group
while retaining as much of the Lie algebraic structure as possible.
Consequently this will allow construction of the gauge theory to proceed
in the standard manner, with the resulting theory being a deformation of the
standard one. Gauge theories based on extensions of simple Lie groups such as
non-semisimple Lie
groups have been considered recently \cite{tseytlin94}. In this letter we will
take a larger step to a theory where the underlying gauge ``group'' is
non-associative. A non-associative algebra has no group structure in the
normal sense but by considering the algebra as a deformation of a Lie algebra
we can obtain the form of the resulting deformation of the gauge field.
For this reason the theory will apparently break the gauge symmetry, but
only when this symmetry is assumed to be of the Lie group form. It is in
this sense that we can regard the resulting theory as one involving
a higher ``non-associative gauge symmetry''.

Our justification for considering non-associativity as the mechanism for
extending the Lie group structure is twofold. Firstly, non-associative
algebras have been linked with a number of interesting gauge groups. The
exceptional GUT groups, such as $E_6$, and the internal symmetry group of the
anomaly free heterotic string $E_8 \times E_8$ have in common the
fact that they are automorphism groups of the non-associative exceptional
Jordan
algebra $M_3^8$ of $3 \times 3$ matrices over the octonions
\cite{jordan33,gursey78,sorgsepp79}. G\"unaydin \& G\"ursey
\cite{gunay73,gunay74} also used the fact
that $SU(3)$ is a subgroup of the automorphism group of the octonions to
obtain a theory of quark confinement, which was subsequently extended by
Dixon \cite{dixon90a,dixon90b}.
Although the gauge groups in these cases are not strictly non-associative
this common link is suggestive of a deeper underlying non-associative
``symmetry''.

The non-associative octonions, the last in the sequence of four division
algebras of the Hurwitz theorem, have also been linked to spacetime
symmetries in 10 dimensions. The Lorentz ``group'' in this case is
essentially $SL(2)$ over the octonions \cite{kugo83,foot87a}.
Consequently octonionic
spinors \cite{kugo83,davies86,chung87,tachi89} have also been linked
to 10 dimensional spacetime and
the Green-Schwarz superstring finds a natural formulation in terms
of the exceptional
Jordan algebra $M_3^8$ \cite{foot86,foot87a,foot87b,foot88,foot89a}.
These correlations, and the association of
supersymmetry in ten dimensions with the octonions
\cite{kugo83,evans88,tachi89},
suggests that a non-associative internal symmetry may be particularly
relevant for theories in ten spacetime dimensions.
Finally, on a more technical point, non-associative
structures such as 3-cocycles
have been linked with chiral anomalies in field theories
\cite{jackiw85a,jackiw85b,hou86} and therefore removal of such problems may
also require a non-associative description.

Our second justification for considering a non-associative deformation,
and as a motivation behind the algebraic structure we shall
choose, is that it provides a framework for considering tensor product gauge
groups, i.e.
$ G  =  A \otimes B \otimes C \otimes \ldots$,
where there is some {\em coupling\/} between the algebras of the different
elements.
The coupling then implies that the gauge group is no longer a direct product
and therefore a more complete unification of the groups $A,B,C,...$ into
the group $G$ is achieved.

The major problem with considering a non-associative gauge theory is that
a gauge group in the normal sense does not exist, due to the non-associativity.
Our technique for dealing with this
owes its inspiration partly to the gauge theories considered by
Waldron \& Joshi \cite{waldron92}, and Lassig \& Joshi \cite{lassig95}, where
the gauge algebra was that of the octonions.
The octonions form a non-associative alternative algebra, and thus a
generalisation of the Lie group approach to gauging is required.
Our approach will not specifically involve octonions but for
clarity it is worth reviewing the form of the octonionic gauge theory making
reference to how it relates to the generalisation of gauge theories.

It is well known that the octonionic units can be
represented in terms of (associative) left and right matrices in the bimodular
representation. The details of how this representation is obtained have been
considered previously (see \cite{sorgsepp79,waldron92,lassig95})
and will not be reproduced here.
In this representation the non-associativity is manifest in the inability of
either the left matrix or right matrix algebra to {\em close\/}.
When considered in isolation the left matrices can be considered as generators
of a Lie algebra where the extra generators required to close the algebra
are missing. In the octonion case these missing generators are replaced by
a coupling between the left and right matrices. i.e. the left matrix algebra
is ``closed'' via a coupling to the right matrices, and vice versa.

If we were to consider only the left matrices as a gauge algebra
\cite{waldron92} then the ``missing'' generators in the Lie algebra, and their
construction via coupling to another algebra (the right matrices), give
the new physics which will become apparent in the resulting gauge field
Lagrangian. This can be made more explicit by noting that the left matrices
in the bimodular octonion representation are also generators of the $SO(8)$
symmetry group. Thus the octonionic symmetry can be visualised as some
particular observable channels of the $SO(8)$ symmetry.
Importantly calculations can still be made as though the full $SO(8)$, i.e. Lie
group, symmetry were present. The restriction on the generators available
will then lead to the new physics in the system.

The octonionic case discussed qualitatively above was used to indicate how the
methodology of using a non-associative generalisation of the Lie algebra
structure allows the Lie structure of the gauge group to be retained modulo
the closure problem. The fact that the standard Lie algebraic calculational
techniques can still be used (cf. QGGTs) implies that we can consider a
standard
Yang-Mills type theory. Having considered this possibility qualitatively
we propose a possible algebraic structure for the gauge symmetry in Section 2,
and develop the corresponding gauge theory in Section 3. A possible
correspondence
between this approach and QGGTs is also considered in this section.

\section{A Parameterised Non-associative Algebra}
Standard Yang-Mills type gauge theories based on internal symmetries
described by non-Abelian Lie groups have been extraordinarily successful
in particle physics with electroweak theory and QCD being the most prominent
examples. Some possible reasons for generalising this structure were
mentioned in Section 1, however it is clear that we would wish to retain
most of the nice features of these theories. Thus we would expect the
generalised theory to reduce to the standard theory in the limit of some
parameter,
as in QGGTs. We can achieve this, and our previously mentioned aim to consider
coupling in tensor product gauge groups, in our non-associative formalism
with the
following algebraic structure of the gauge group.

Consider $M$ sets of $N$ generators which we can represent in the following
matrix.
\be
 (T)_{ji} & = & \left(\begin{array}{c}
		    T_{1i} \\
		    T_{2i} \\
		    \vdots \\
		    T_{Mi}
                \end{array}\right)
         = \left( \begin{array}{cccc}
		     T_{11}  &          &    \cdots     & T_{1N} \\
		     T_{21}  & T_{22}   &               &    \\
		     \vdots  &          &   \ddots      & \vdots   \\
		     T_{M1}  &          &   \cdots      & T_{MN}
		  \end{array}   \right).
\ee
For generality we allow the possibility that the algebras are not all of equal
dimension. Then for a particular set of generators it may be that
\be
 T_{(p)(1)}\ldots T_{(p)(r)} \neq 0 & \;\;\;\;\; &
      T_{(p)(r+1)}\ldots T_{(p)(N)} = 0.
\ee
We regard each set of generators $T_{pi}$, for fixed $p \in 1..M$, as the
generators of a simple Lie algebra which may or
may not {\em close\/}. We
represent this in the following way:
\be
 [T^p_i,T^p_j] & = & f_{ijk}^p T_k^p + \sum^M_{n=1} \si^p_n [T^n_j,T^p_i],
\ee
where $p \in 1..M$ and the $\si^p_n$ are constants. In this representation
the parameters $\si^p_n$, for $n \in 1..M$, determine the closure of each set
of generators $T_{pi}$. For a given $p \in 1..M$, if $\si_n^p = 0\;\; \forall
n$
then the generators $T_{pi}$ close and we have a normal associative Lie
algebra.
If, however, there exist $n \in 1..M, \;\; n \neq p$ such that $\si_n^p \neq 0$
then the generators $T_{pi}$ do not close and this is manifest in some
{\em mixing\/} between the sets of generators. This {\em nonlinearity\/}
will imply non-associativity for the algebra of the set $T_{pi}$.

This can be made explicit by considering the {\em Jacobi function\/}
\be
 J (a,b,c) & = & [a,[b,c]]+[b,[c,a]]+[c,[a,b]],
\ee
where $J(a,b,c)=0$ for $a,b,c$ elements of an associative algebra.
For the case at hand, for fixed $p$,
\be
 J(T^p_i,T^p_j,T^p_k) & = & \si_n^p \left([T_i^p,[T_k^n,T_j^p]] +
      [T_j^p,[T_i^n,T_k^p]]+[T_k^p,[T_j^n,T_i^p]] \right).
\ee
Thus associativity is restored for $\si_n^p=0 \;\; \forall n \in
1..M$ or alternatively if all the sets of generators commute. We can
write this explicitly in terms of the {\em associator\/},
\be
 (a,b,c) & = & (ab)c - a(bc),
\ee
by noting
\be
 \ep^{ijk}(a_i,a_j,a_k) & = & J(a_i,a_j,a_k).
\ee

Thus we have an algebraic structure where the non-associativity is manifest
in the inability of the set of Lie algebraic generators $T_{pi}$ to close.
This ensures that the generators will still have a matrix representation, and
we can retain the nice features of Lie algebras, and to some extent its group
structure. The non-associativity can be ``turned on'' by closing the algebra
via mixing with generators from other sets.

\section{Non-associative Gauging}
We consider a Yang-Mills type theory where the gauge ``group'' has the
algebraic structure ${\cal A}$ considered in the previous section:
\be
 [T_i^p,T_j^p] = f_{ijk}^p T_k^p + \si_n^p [T_j^n,T_i^p],
\ee
where the $\si_n^p$ are constants, and the summation over $n=1..M$ is implicit.
We assume $p \in [1..M]$ and that the indices $i,j,k \in [1..N]$ label the
elements of each particular set of generators.
The constants $\si_n^p$ parameterise the level of non-associativity.

The Lie algebraic structure retained in this algebra allows the Yang-Mills
gauge theory to be developed in the standard way.
We introduce the following matter fields:
\be
 \ps & = & \left( \begin{array}{c}
		   \ps_1 \\
		   \ps_2 \\
		   \vdots \\
		   \ps_D
		  \end{array}\right),
\ee
where $D$ is the dimension of the representation of ${\cal A}$.
Since the constituent generators of ${\cal A}$ are equivalent to those of a
Lie algebra we can consider exponentiating to produce a structure which,
in the limit $\si_n^p \rightarrow 0 \;\; \forall n$, will correspond to a
Lie group.
This allows consideration of the following standard gauge transformation,
\be
 \ps(x) \longrightarrow \ps '(x) & = & e^{-i\ta_i^p \theta_p^i(x)} \ps(x)
       = U(\theta_p) \ps(x),
\ee
where $p \in 1..M$ is not summed. The particular representation of
${\cal A}$, in
this case given by $D \times D$ matrices $\ta_i^p$, satisfies the algebra,
\be
 [\ta_i^p,\ta_j^p] & = & f_{ijk}^{p'} \ta_k^p + \si_n^{p'} [\ta_j^n,\ta_i^p],
     \label{alg}
\ee
and we impose
\be
 \mbox{Tr} (\ta_i^p \ta_j^p) & = & \frac{1}{2} \de_{ij} \;\;\;\;\;\;\;\;\;
	 \forall p \in 1..M,
\ee
on the adjoint representation via
the relevant normalisation. Since we will henceforth work only with this
representation the primes in $f_{ijk}^{p'}$ and $\si_n^{p'}$ will be
suppressed.

We can now proceed in the standard way by defining the covariant derivative
as
\be
 {\cal D}_{\mu}^p & = & \ptl_{\mu} + ig A_{\mu}^p(x), \label{covd}
\ee
where
\be
 A_{\mu}^p(x) & = &  ^i A_{\mu}^p(x) \ta_i^p, \label{gfields}
\ee
and summation over $i = 1..N$, but not $p \in 1..M$, is implicit. This
explicitly introduces the vector gauge fields $^i A_{\mu}^p(x)$ which,
in the standard case, would ensure
that the Lagrangian density ${\cal L}$ is invariant under local gauge
transformations.
The $A_{\mu}^p(x)$ transform as
\be
 A_{\mu}^p \longrightarrow A_{\mu}^{p'} & = & U A_{\mu}^p U^{-1} -
       \frac{i}{g} U \ptl_{\mu} U^{-1}.
\ee
On evaluation for infinitesimal transformations,
\be
 A_{\mu}^{p'} & = &  \left(^i A_{\mu}^{p'} \right)\ta_i^p +
      i\si_n^p \theta_p^i(x)\; ^j A_{\mu}^p [\ta_j^n,\ta_i^p],
\ee
where
\be
 ^i A_{\mu}^{p'} & = & ^i A_{\mu}^p + f_{ijk}^{p'} \theta_p^i\;  {^k}A_{\mu}^p
	   + \frac{1}{g} \ptl_{\mu} \theta_p^i (x),
\ee
is the algebraically closed part in the usual form.
Thus the non-associativity is manifest in the
inability of the transformation
to close. We note that for $\si_n^p=0\;\; \forall n$ the
transformation does close as required.

The antisymmetric curvature tensor can be defined in the normal way:
\be
 F_{\mu\nu}^p & = & -\frac{i}{g} [{\cal D}_{\mu}^p,{\cal D}_{\nu}^p]
       \nonumber \\
          & = & \ptl_{[\mu}A^p_{\nu ]} + ig [A_{\mu}^p,A_{\nu}^p]
	       \nonumber \\
          & = & ^iF_{\mu\nu}^p \ta_i^p + i\si_n^p g\;
		 ^jA_{\mu}^p \;{^k} A_{\nu}^p
		       [\ta_k^n,\ta_j^p],
\ee
where again,
\be
 ^iF_{\mu\nu}^p & = & \ptl_{[\mu} ^i A_{\nu]}^p - g f_{ijk}^p \;{^j}A_{\mu}^p
      \;{^k}A_{\nu}^p,   \label{curvti}
\ee
is the algebraically closed part.

We can now evaluate the gauge field kinetic term in the Lagrangian density. We
obtain:
\be
 {\cal L}_{gauge}^p & = & - \frac{1}{2} \mbox{Tr} (F_{\mu\nu}^p F^{p\mu\nu})
	\nonumber \\
	    & = & -\frac{1}{2}\; ^iF_{\mu\nu}^p\; ^iF^{p\mu\nu} -
		  \frac{1}{2}i\si_n^p g \;^jA_{\mu}^p \; ^kA_{\nu}^p \;
          	      ^tF_{\mu\nu}^p (\mbox{Tr}([\ta_k^n,\ta_j^p]\ta_t^p))
			      \nonumber\\
            &   & -\frac{1}{2}i\si_n^p g \;^uA_{\mu}^p \; ^vA_{\nu}^p \;
		     ^iF_{\mu\nu}^p (\mbox{Tr}(\ta_i^p[\ta_v^n,\ta_u^p]))
			      \nonumber\\
            &   & +\frac{1}{2}(\si_n^p)^2 g^2 \;^jA_{\mu}^p \; ^kA_{\nu}^p \;
			  ^uA_{\mu}^p \; ^vA_{\nu}^p
			      (\mbox{Tr}([\ta_k^n,\ta_j^p][\ta_v^n,\ta_u^p])).
\ee

This Lagrangian represents the kinetic term for a gauge field where the algebra
of the primary generators (the p'th set) is altered by mixing with external
generators. We note that the terms corresponding to this mixing are suppressed
by factors of $\si_n^p$, and thus by setting these coupling constants to zero
the non-associativity is turned off and a standard Yang-Mills gauge kinetic
term
results.

We note that this Lagrangian is however
biased in favour of the p'th set of generators,
with the other sets entering via the nonlinear algebraic relations.
This is the relevant situation if we are considering a small coupling
between a primary algebra and a secondary algebra, however the implicit bias
towards the $p^{th}$ set used so far may be artificial in others.
We can obtain a symmetric Lagrangian density for the gauge field by
summing the contributions $\forall p \in 1..M$. This implies the gauge
transformation is now
\be
 \ps(x) \rightarrow \ps '(x) & = & e^{-i\ta_i^p \theta_p^i(x)} \ps(x)
\ee
where now both $i$ and $p$ are summed. The relations for the covariant
derivative, Eq. \ref{covd}, and the gauge fields, Eq. \ref{gfields}, can now
be reinterpreted with $p$ summed over $1..M$. The
covariant derivative now takes on a form similar in appearance to that
encountered with tensor product gauge groups. The difference being in that
here there exists the possibility for coupling between the components.
The curvature tensor in
symmetrised form is then,
\be
 F_{\mu\nu} & = & \ptl_{\mu}A_{\nu}^p - \ptl_{\nu}A_{\mu}^p +
       ig[A_{\mu}^p,A_{\nu}^p] + ig[A_{\mu}^r,A_{\nu}^s],
\ee
where $p,r,s = 1..M,\;\; r \neq s$. Thus we have
\be
 F_{\mu\nu} & = & ^iF_{\mu\nu}^p \ta_i^p + ig\si_n^p \;{^j}A_{\mu}^p
	 \;{^k}A_{\nu}^p
      [\ta_k^n,\ta_j^p] + ig \;^tA_{\mu}^r \;{^u}A_{\nu}^s [\ta_t^r,\ta^s_u],
\ee
where $^iF_{\mu\nu}^p$ is given by Eq. \ref{curvti}, $i,j,k,t,u$ are summed
over $1..N$, and $p$ is now summed over $1..M$.
This symmetric formulation fundamentally alters the covariant
derivative and thus the limit $\si_n^p \rightarrow 0$ alone no
longer reduces the
theory to one with no coupling. If, however, all the generators in different
sets commute then the Lagrangian for each set will decouple and have
the standard form.

This will obviously lead to a more complicated gauge kinetic term for the
symmetrised Lagrangian which we include for completeness,
\be
 {\cal L}_{gauge} & = & - \frac{1}{2} \mbox{Tr} (F_{\mu\nu} F^{\mu\nu})
	\nonumber \\
	    & = & -\frac{1}{2}\; ^aF_{\mu\nu}^p \; ^f F_{\mu\nu}^q \mbox{Tr}
			      [\ta_a^p\ta_f^q]
		    - \frac{1}{2}i\si_m^q g \;^gA_{\mu}^q \; ^hA_{\nu}^q \;
			^a F_{\mu\nu}^p \mbox{Tr}[\ta_a^p[\ta_h^m,\ta_g^q]]
			 \nonumber\\
            &   &   -\frac{1}{2}i g \;^iA_{\mu}^t \; ^jA_{\nu}^u \;
			^a F_{\mu\nu}^p \mbox{Tr}[\ta_a^p[\ta_i^t,\ta_j^u]]
	           -i\si_n^p g \; ^b A^p_{\mu} \; ^c A_{\nu}^p \; ^fF_{\mu\nu}^q
			\mbox{Tr}[[\ta_c^n,\ta_b^p]\ta_f^q]
			   \nonumber\\
            &   &  + \frac{1}{2}\si_n^p \si_m^q g^2 \; ^b A_{\mu}^p \;
		  ^c A_{\nu}^p \;^gA_{\mu}^q \; ^hA_{\nu}^q \;
			     \mbox{Tr}[[\ta_c^n,\ta_b^p][\ta_h^m,\ta_g^q]]
			       \nonumber\\
            &   &  + \frac{1}{2}\si_n^p g^2 \; ^bA_{\mu}^p \;
			  ^cA_{\nu}^p \;^iA_{\mu}^t
		       \; ^jA_{\nu}^u \;
			     \mbox{Tr}[[\ta_c^n,\ta_b^p][\ta_i^t,\ta_j^u]]
			      \nonumber\\
	    &   & -\frac{1}{2} i g \; ^dA_{\mu}^r \; ^eA_{\nu}^s \;
		      ^fF_{\mu\nu}^q \; \mbox{Tr}[[\ta_d^r,\ta_e^s]\ta_f^q]
			   \nonumber\\
            &   & + \frac{1}{2} \si_m^q g^2 \; ^dA_{\mu}^r \;
		    ^eA_{\nu}^s \;^gA_{\mu}^q
		    \; ^hA_{\nu}^q \; \mbox{Tr}[[\ta_d^r,\ta_e^s]
			      [\ta_h^m,\ta_g^q]] \nonumber\\
            &   & + \frac{1}{2} g^2 \; ^dA_{\mu}^r \; ^eA_{\nu}^s
		       \;^iA_{\mu}^t \;
		    ^jA_{\nu}^u \;\mbox{Tr}[[\ta_d^r,\ta_e^s]
			      [\ta_i^t,\ta_j^u]]
\ee
where $a,b,c,d,e,f,g,h,i,j = 1..N$ and $p,q,r,s,t,u=1..M$,
$r \neq s$, $t \neq u$.
This gives us the full gauge field kinetic term in the general case, which
represents the result of the non-associative deformation of the gauge group.
With regard to renormalisation, superficially
this Lagrangian density has no terms of higher than quartic
power in the gauge fields. However a full
consideration of renormalisability would
require calculations to loop level which have not been considered as this
is very much a toy model at this stage.

The analysis has been of a general nature thus far. We will now indicate
how the general algebraic structure introduced in Section 2 includes
various subalgebras which have been previously been
considered as possible gauge algebras.

\subsection{Specific Cases}

\subsubsection{Associative Lie Algebras}
The standard form gauge groups are realised trivially when
\be
 \si_n^p & = & 0\;\;\; \forall n \in 1..M.
\ee
The generator sets then decouple and there is no need for symmetrisation.
The normal Yang-Mills type Lagrangian density results. i.e.
\be
 {\cal L}_{gauge} & = & - \frac{1}{4} F^i_{\mu\nu}F^{i\mu\nu}.
\ee
In practice since there are so many nice features of these theories, we might
expect that the situation of most interest in the low energy regime
might be when the $\si_n^p$
are small,
and thus the theory approaches the Lie algebra case. The Lie algebra structure
of the gauge group would then only be slightly perturbed by the induced
non-associativity.

\subsubsection{Octonionic Algebras}
Gauge theories based on an octonionic gauge algebra have been considered
before \cite{waldron92,lassig95} and they represent the first non-trivial
instance of the general algebraic structure considered. This is observed by
using the left/right matrix bi-representation for octonions. This implies two
sets of generators, the left and right matrices in this case. The gauge
theories mentioned above, in the notation of Lassig \& Joshi \cite{lassig95},
can be realised when
\be
 \si_n^p & = & \left\{ \begin{array}{ll}
	 2    & \mbox{for}\;\; n,p=1,2;\;\; n \neq p \\
	 0    & \mbox{otherwise}
		     \end{array}\right. ,
\ee
and
\be
 f_{ijk}^p & = & (-1)^{p-1} \frac{i}{2} \ep_{ijk},
\ee
where $\ep_{ijk}$ is the anti-symmetric tensor for octonions where, using the
standard multiplication table,
$\ep_{ijk}=1$ for each cycle $ijk=123,145,176,246,257,347,365$.
Thus there are two coupled sets
of generators, which are the $\la_i,\rh_i$ matrices of \cite{waldron92} and
\cite{lassig95}, representing the left and right matrices of the bimodular
representation.

\subsubsection{Quantum Groups}
As mentioned in the introduction quantum group gauge theories have been the
subject of considerable recent interest. In these theories the process
of gauging was altered to generate a more general quantum group
symmetry. In contrast, in this discussion we have retained
the standard Yang-Mills machinery allowing the standard symmetry to be broken
by the deformation. Despite this difference in approach we show in this
section how the general algebraic structure can be linked to the
quantum universal enveloping algebra of a Lie algebra.

For concreteness we consider the quantum universal enveloping algebra
$U_q(su(2))$ in the Drinfel'd-Jimbo basis (see for
example \cite{biedenharn89,macfar89}), whose
generators satisfy the following relations
\be
 [J_{\pm},J_3] & = & \mp J_{\pm}\;\;\;\;\;\;\;\;\; [J_{+},J_{-}] = [2J_3]_q,
\ee
where, in the notation of Macfarlane \cite{macfar89},
the $q$-integers are given by
\be
 [x]_q & = & \frac{q^{x}-q^{-x}}{q-q^{-1}}.
\ee
Writing $J_{\pm} = J_1 \pm iJ_2$ we have
\be
 \left[J_1,J_2\right] & = & \frac{1}{2}i [2J_3]_q \nonumber\\
 \left[J_2,J_3\right] & = & \frac{1}{2}i (2J_1) \label{q-alg} \\
 \left[J_3,J_1\right] & = & \frac{1}{2}i (2J_2). \nonumber
\ee
To make this tractable, we assume $q=1+\de$, where $|\de| \ll 1$, but
$\de \neq 0$. Then we
can expand the $q$-integers as a power series and we have
\be
 q^{2J_3}-q^{-2J_3} & = & (1+\de)^{2J_3}-(1+\de)^{-2J_3} \nonumber \\
      & \approx & \sum_{n=1}^{\infty} C_n J_3^n,
\ee
where
\be
 C_n & = & C_n (\de^n, \de^{n+1}, \de^{n+2}, \ldots)
\ee
is a constant, which will be convergent for small $\de$. Note that in this case
the above approximation becomes exact. As an aside
we also note that for $U_q(su(2))$
$C_n$ is only non-zero for odd n.

In this case we can now rewrite Eq.s \ref{q-alg} as
\be
 &  &
\begin{array}{lll}
 \left[J_1,J_2\right] & = & i \left( \frac{C_1}{2(q-q^{-1})}\right)J_3 +
	\frac{i}{2(q-q^{-1})} \sum_{n=2}^{\infty} C_n J_3^n \\
 \left[J_2,J_3\right] & = & iJ_1 \\
 \left[J_3,J_1\right] & = & iJ_2.
\end{array}
\ee
Thus the algebra can be represented in the following form
\be
 [J_i,J_j] & = & f_{ijk}J_k + N_{ijk},
\ee
with the nonzero antisymmetric structure constants
\be
 f_{ijk} & = & i\ep_{ijk} + i\ep_{ijk}\de_{k3}\left( \frac{C_1}{2(q-q^{-1})}
	   -1 \right),
\ee
where $\ep_{ijk}$ is the Levi-Civita antisymmetric tensor, and the
extra term $N_{ijk}$, which represents the inability of the algebra to close,
is given by
\be
 N_{ijk} & = &
      i\ep_{ijk} \de_{k3} \frac{1}{2(q-q^{-1})} \sum_{n=2}^{\infty} C_n J_3^n.
\ee
We now assume that there exist generators $K_{ni}$, where $n \in 2..\infty$
and $i,j \in 1,2,3$, such that
\be
 [K_{nj},J_i] & = & \ep_{ijk} \de_{k3} J_k^n .
\ee
Thus we  have
\be
 N_{ijk} & = & \frac{i}{2(q-q^{-1})} \sum_{n=2}^{\infty} C_n [K_{nj},J_i].
\ee

Therefore we can finally represent the algebra as
\be
 [J_i,J_j] & = & f_{ijk} J_k + \frac{i}{2(q-q^{-1})} C_n [K_{nj},J_i],
\ee
where the summation over $n=2..\infty$ is implicit. This is exactly the form
required for our general procedure, and thus the $q$-algebra is realised in
this
form when we interpret
\be
 M & = & \infty \nonumber \\
 \si_n & = & \frac{i}{2(q-q^{-1})} C_n.
\ee
It is therefore apparent that to fully represent the quantum group as a gauge
group in this manner requires mixing between an infinite number of generators!
We should also note that the algebra is not symmetric and the required gauge
term in the Lagrangian would also be unsymmetric with respect to all the
generators. This is due to the bias towards the set $\{J_i\}$ as a result of
our construction.

Clearly it would appear that this formalism is not particularly useful
for considering quantum gauge groups, however for practical calculational
purposes we can note that if
$q \sim 1$ then the terms in the infinite series for $C_n$ would quickly
tend to zero. Then a truncation of the series would seem reasonable and the
constants $\si_n$ could be explicitly evaluated giving, in effect, a
perturbation to a given order in $(1-q)$. The perturbation series would
then interpolate to some extent between the Lie algebra and the full
quantum group, for the case when $q \sim 1$.

\section{Concluding Remarks}
In this letter we have considered a possible formalism for the
analysis of a gauge
theory based on a ``group'' with a non-associative algebraic structure.
The gauge algebra presented was obtained initially
as a generalisation of the standard Lie algebraic structure, where coupling
between different Lie algebras is allowed. This non-associative formalism is
apparently quite different from other Lie algebra generalisations such as
the infinite dimensional Kac-Moody algebras.

Gaining inspiration from the bimodular
representation of octonions the
non-associativity inherent in the algebra, for ease of calculation,
is implicit as a
closure problem for the algebra of the generators.
The analysis of the resulting gauge theory has been somewhat cursory, and
in particular renormalisability for this toy model has not been considered.

Finally, we would like to point out that the viewpoint taken
in this letter has resulted in a theory
where the gauge symmetry is broken via the non-associative coupling between
the sets of generators. This formalism was used to explicitly show the
new physics obtained by deforming the symmetry, and this explicit gauge
symmetry
breaking is manifest when the constants $\si^p_n$ become non-zero and the
individual generator sets no longer close. An alternative viewpoint is that
although a Lie group symmetry has been broken a higher ``non-associative''
symmetry is retained. This viewpoint requires a significant alteration
to the current understanding of what constitutes a gauge symmetry. If this
seems to radical then an alternative to our construction could be
considered where a full (generalised) symmetry is retained. This is the
approach taken with QGGTs and would require
deformation of the standard gauge transformations,
and most likely the form of the gauge field Lagrangian density.
The Lie group type symmetry is then
deformed and reduces to the standard one in the limit of some parameter.
It is clear
that obtaining such a theory where a full symmetry, in something
approaching the Lie group sense,
is retained
would be worthwhile, and this is under investigation.

\end{document}